\documentclass[prl,twocolumn,superscriptaddress]{revtex4-1}

\usepackage{amsmath,amssymb}
\usepackage{graphicx}
\usepackage{epstopdf}
\usepackage{bm}
\usepackage{color}

\newcommand{\as}	{$^{75}$As}
\newcommand{\tnem}	{$T_\text{nem}$}

\newcommand{\slr} 	{$T_1^{-1}$}
\newcommand{\slrt} 	{$(T_1T)^{-1}$}

\newcommand{\splitting} 	{$\Delta\nu_{\parallel a}$}

\begin{document}

\title{Nematicity and magnetism in LaFeAsO single crystals probed by $^{75}$As nuclear magnetic resonance} 

\author{J. M. Ok}
\affiliation{Department of Physics, Pohang University of Science and
Technology, Pohang 790-784, Korea}
\affiliation{Center for Artificial Low Dimensional Electronic Systems, Institute of Basic Science, Pohang 790-784, Korea} 
\author{S.-H. Baek}
\email[Corresponding author: ]{sbaek.fu@gmail.com}
\affiliation{IFW Dresden, Helmholtzstr. 20, 01069 Dresden, Germany}
\author{D. V. Efremov} 
\affiliation{IFW Dresden, Helmholtzstr. 20, 01069 Dresden, Germany}
\author{R. Kappenberger}
\affiliation{IFW Dresden, Helmholtzstr. 20, 01069 Dresden, Germany}
\author{S. Aswartham}
\affiliation{IFW Dresden, Helmholtzstr. 20, 01069 Dresden, Germany}
\author{J. S. Kim}
\affiliation{Department of Physics, Pohang University of Science and
	Technology, Pohang 790-784, Korea}
\affiliation{Center for Artificial Low Dimensional Electronic Systems, Institute of Basic Science, Pohang 790-784, Korea}
\author{Jeroen van den Brink}
\affiliation{IFW Dresden, Helmholtzstr. 20, 01069 Dresden, Germany}
\affiliation{Department of Physics, Technische Universit\"at Dresden, 01062 Dresden, Germany}
\author{B. B\"uchner}
\affiliation{IFW Dresden, Helmholtzstr. 20, 01069 Dresden, Germany}
\affiliation{Department of Physics, Technische Universit\"at Dresden, 01062 Dresden, Germany}

\date{\today}

\begin{abstract}
	We report a $^{75}$As nuclear magnetic resonance study in LaFeAsO single 
	crystals, which undergoes nematic and  
	antiferromagnetic transitions at $T_\text{nem}\sim 156$ K and $T_N 
	\sim 138$ K, respectively.  Below \tnem, the \as\  
	spectrum splits sharply into two for an 
	external magnetic field parallel to the orthorhombic $a$ or $b$ axis in the FeAs planes. Our analysis of the data demonstrates that the NMR line splitting arises from an electronically driven rotational symmetry breaking. The \as\ spin-lattice relaxation rate as a function of temperature shows that spin  
	fluctuations are strongly enhanced just below \tnem. These NMR findings 
	indicate that nematic order promotes spin fluctuations in magnetically ordered LaFeAsO, as observed in non-magnetic and superconducting FeSe. We conclude that the origin of nematicity is identical in both FeSe and LaFeAsO regardless of whether or not a long range magnetic order develops in the nematic state. 
\end{abstract}

\pacs{}

\maketitle

Understanding nematic order and its relationships to magnetism and 
superconductivity remain among the most important questions in the current 
study of Fe-based superconductors (FeSCs) \cite{fernandes14,bohmer15a,si16,kuo16,yamakawa16}. While different classes of FeSCs show very similar softening of the lattice and divergence of  
the nematic susceptibility when approaching the nematic transition \cite{fernandes13,bohmer15}, there are 
pronounced differences with respect to the slowing down of spin fluctuations (SFs). 
In the BaFe$_2$As$_2$-type system, the dynamic spin susceptibility as revealed from 
NMR spin-lattice relaxation rate data scales with the softening of the elastic constant 
above the nematic transition \cite{fernandes13}. This was interpreted as 
evidence for theoretical scenarios where nematic order is driven by an 
antiferromagnetic instability. On the other hand, the nonmagnetic compound FeSe does 
not show any slowing down of SFs above the nematic transition 
\cite{baek15,baek16,bohmer15,wang15a}. 
This, as well as the absence of long-range magnetic order in FeSe, has been 
taken as evidence for an alternative origin of nematic order, related to 
orbital degrees of freedom \cite{onari16,yamakawa16}. 
Based on a renormalization group (RG) analysis  
\cite{chubukov16,classen17}, a possible scenario for such  
an orbital order as the leading instability was derived. According to this model 
the origin of the orbital order in FeSe is the small Fermi energies of the electron and hole bands. While the changes of the electronic structure 
due to nematic order obtained by the RG analysis are consistent with recent high 
resolution angle-resolved photoemission spectroscopy (ARPES) studies \cite{watson15,suzuki15,fanfarillo16}, it is impossible to 
prove this scenario from the  
available data on FeSe. Several alternative scenarios were suggested for the 
nematic order in FeSe, for example, related to the frustration of the
magnetic exchange interactions \cite{glasbrenner15, wang15}.

In order to shed more light on the possible origin of nematic order in 
FeSCs, we carried out a NMR investigation on a prototypical FeSC, LaFeAsO, which shows a clear separation between nematic order at 
$T_\text{nem}\sim 156$ K and magnetic order at $T_N \sim 138$ K. The temperature dependence 
of \as\ NMR spectra  
and spin-lattice relaxation rates measured in our LaFeAsO single crystals reveals remarkable similarities to that in FeSe, 
suggesting that the scaling behavior found in BaFe$_2$As$_2$ is not generic for 
FeSCs. In contrast, qualitatively the interplay between 
nematicity and magnetism as seen by NMR is almost identical in FeSe and LaFeAsO. 
The only difference is that, in the latter, the impact of orbital order on SFs
 is much stronger and the pronounced slowing down of SFs is 
followed by long range magnetic order 15 K below \tnem. 

Single crystals of LaFeAsO were grown by using NaAs-flux techniques. The mixture of LaAs, Fe, Fe$_2$O$_3$ and NaAs powders with a stoichiometry of LaAs:Fe:Fe$_2$O$_3$:NaAs = 3:1:1:4 was double-sealed with a Ta tube (or stainless steel tube) and an evacuated quartz tube. The entire assembly was heated to 1150$^\circ$C, held at this temperature for 40 h, cooled slowly to 700$^\circ$C at a rate of 1.5$^\circ$C/h and then furnace-cooled. The NaAs flux was rinsed off with deionized water in a fume hood and the plate-shaped single crystals were mechanically extracted from the remaining by-products. 

\begin{figure*}
	\centering
	\includegraphics[width=0.75\linewidth]{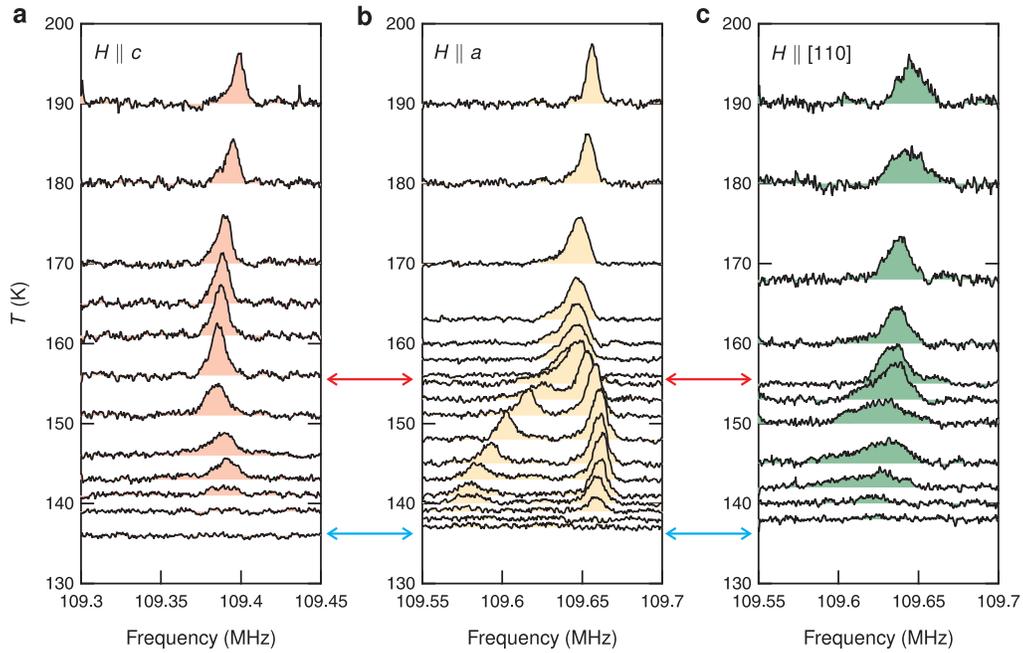}
	\caption{ \as\ NMR spectra measured at 9 T applied parallel to 
		the $c$ axis (a), to the $a$ axis (b), and to the [110] 
		direction (c), as a function of temperature. The red and blue horizontal arrows denote the nematic and the magnetic transition temperatures, \tnem\ and $T_N$, respectively. The  
		\as\ line splits below \tnem\  for $H\parallel a(b)$, but it 
		remains a single line for $H\parallel c$ 
		and $H\parallel [110]$, proving spontaneous $C_4\rightarrow C_2$ symmetry breaking. 
		The notable broadening below \tnem\ for $H\parallel [110]$ 
		is attributed to a slight misalignment of the sample.}
	\label{fig_spec}
\end{figure*}

\as\ (nuclear spin $I=3/2$) NMR measurements were carried out on LaFeAsO single crystals at an 
external field of 9 T and in the range of temperature 130 -- 300 K.  It turned out 
that the \as\ NMR spectrum becomes significantly narrower for smaller crystals, 
indicating that local inhomogeneity or disorder increases rapidly in proportion to the size of the crystal.
Since the coupling of \as\ nuclei to nematicity is generally very small in FeSCs,  the sufficiently narrow NMR spectrum is crucial for  
the detailed investigation of nematicity. For this reason,  we collected and aligned five single crystals
as small as 0.15$\times$0.15$\times$0.01 mm$^3$ to achieve a measurable signal 
intensity while maintaining a minimal linewidth. The 
alignment of samples is satisfactory, based on the much narrower \as\ line than 
that observed in a previous NMR study \cite{fu12} (see Fig. \ref{fig_spec}). 
The aligned single crystals were reoriented using a
goniometer for the accurate alignment along the external field. The \as\ NMR
spectra were acquired by a standard spin-echo technique with a typical $\pi/2$
pulse length 2--3 $\mu$s.
For the nuclear spin-lattice relaxation (\slr) measurements, we used a large single crystal with 
a dimension of $2 \times 2 \times 0.1$ mm$^3$ \cite{kappenberger18} as inhomogeneity does not 
affect the average spin-lattice relaxation rate.  \slr\ was obtained by fitting the
recovery of nuclear magnetization $M(t)$ after a saturating pulse to
following fitting function,
$$
1-\frac{M(t)}{M(\infty)}=A\left(0.9e^{-6t/T_1}+0.1e^{-t/T_1}\right)$$
where $A$ is a fitting parameter.

Figure \ref{fig_spec} shows the \as\ spectrum as a function of temperature for 
three different field orientations along the $c$, $a$ (or $b$), and [110] 
directions, respectively. The full width at half 
maximum (FWHM) of the line remains very narrow (less than  
20 kHz), evidencing a high quality of the samples. Below \tnem\ we observed a 
clear splitting of the \as\ line for $H\parallel a$. In strong 
contrast, the \as\ line for $H\parallel c$ remains a single line until it 
disappears due to the antiferromagnetic (AFM) ordering at $T_N$. 
We also confirmed that the \as\ 
line does not split when $H$ is applied parallel to the $ab$ plane in 
the diagonal direction ($H\parallel [110]$). 
Therefore one can conclude that the split lines for $H\parallel a$ arise from the two fully 
twinned nematic domains in the orthorhombic structural phase.    

\begin{figure}
	\centering
	\includegraphics[width=0.75\linewidth]{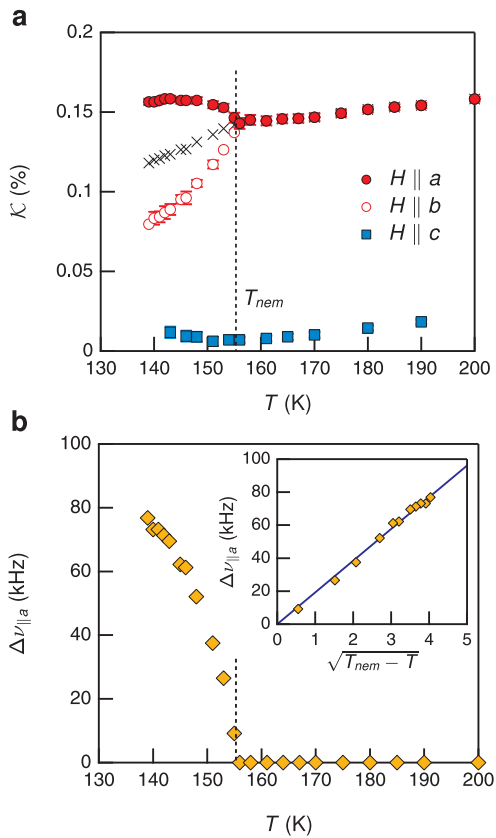}
	\caption{
		(a) Temperature dependence of the \as\ NMR
		shift $\mathcal{K}$ for fields parallel to the $a$ (or $b$) 
		and $c$ axes. The data for $H\parallel ab$ were offset by $-0.1$ for a better comparison. At $T_\text{nem}$, $\mathcal{K}_{\parallel a}$ sharply splits  
		into two, but $\mathcal{K}_{\parallel c}$ shows a smooth $T$-dependence. 
		(b) The \as\ line splitting \splitting\  as a function of temperature. 
		Inset:  \splitting\ is well described by the relation, $\sqrt{T_\text{nem}-T}$ below \tnem, as is  
		expected for an order parameter at a second order phase transition. } 
	\label{fig_knight}
\end{figure}

The temperature dependence of the resonance 
frequency $\nu$ for each NMR lines is presented in Fig. \ref{fig_knight}(a) in 
terms of the NMR  
shift $\mathcal{K}\equiv (\nu-\nu_0)/\nu_0\times 100$ \% where $\nu_0$ is the unshifted 
resonance frequency. 
In a paramagnetic state, the NMR shift can be written as
\begin{equation}
	\label{knight}
\mathcal{K} = A_\text{hf} \chi_\text{spin}+\mathcal{K}_0+\mathcal{K}_\text{quad},
\end{equation}
where 
$A_\text{hf}$ is the hyperfine coupling constant, $\chi_\text{spin}$ the local 
spin susceptibility, $\mathcal{K}_0$ the temperature independent term, and 
$\mathcal{K}_\text{quad}$ the second order quadrupole shift.  
For $T>T_\text{nem}$, the NMR shift for both field directions is weakly 
temperature dependent. Note that the large anisotropy of the NMR shift between the field orientations along $a$ and $c$ is accounted for by the term $\mathcal{K}_\text{quad}$ which is the largest for 
$H\parallel a(b)$, but vanishes for $H\parallel c$. 
The data reveal that the line splitting occurs at \tnem\  and increases upon lowering temperature.  The separation of the two lines, \splitting, exhibits the $\sqrt{T_\text{nem}-T}$  
behavior of a Landau-type order parameter below \tnem, as shown in Fig. \ref{fig_knight}(b). 
This 
indicates that \splitting\ represents the $C_4\rightarrow C_2$ symmetry breaking, or the nematic order parameter.   These features near \tnem\ are identical to the case of FeSe, except for the presence of the AFM transition at  $T_N\sim138$ K.

\begin{figure}
	\centering
	\includegraphics[width=0.7\linewidth]{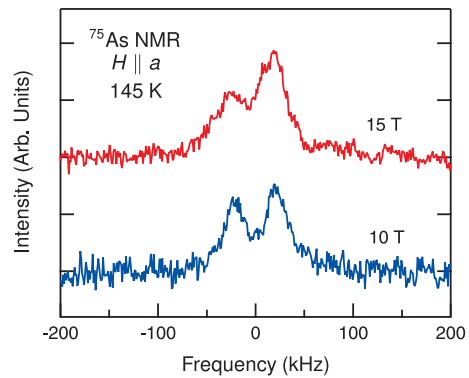}
	\caption{Comparison of the split \as\ spectrum in the nematic phase (at 145 K) for 
		$H\parallel a$ at two different field strengths.  Clearly, the distance between the two peaks does not decrease at a larger field, proving that the \as\ 
		line splitting is not due to a quadrupole effect. Note that the spectrum is considerably broader than those shown in Fig. \ref{fig_spec} because it was obtained in a much larger single crystal used for the spin-lattice relaxation measurements.      
	}
	\label{spec_comp}
\end{figure}

The immediate question then arises which degree of freedom, among lattice/spin/orbital, is 
responsible for the splitting of the \as\ line.    
A previous NMR study in LaFeAsO single crystals \cite{fu12} interpreted that the \as\ line splitting is 
a direct consequence of the quadrupole effect in the twinned orthorhombic 
domains --- that is, the direction of the principal axis of the 
electric field gradient (EFG) in one domain is rotated by 90$^\circ$ in the 
other, giving rise to the different second order quadrupole shift in the two 
domains. 
If this is the case, the separation between \as\ split lines should be given by \cite{bennet}
\begin{equation}
	\label{quad}
	\Delta\nu_{\parallel a}^\text{quad}=\frac{\eta\nu_Q^2}{4\gamma_n H},
\end{equation}
where  
$\eta\equiv |V_{xx}-V_{yy}|/V_{zz}$ is the asymmetry parameter, 
$\nu_Q$ is the quadrupole frequency, and $\gamma_n$ is the nuclear gyromagnetic ratio. 
Accordingly, the line splitting should be inversely proportional to the external field $H$. As shown in Fig. \ref{spec_comp}, however, we  verified that the splitting does not decrease linearly in field, but even slightly increases, as $H$ is increased 
from 10 to 15 T. This unambiguously proves that the line
splitting cannot be ascribed simply to the quadrupole effect. Rather, similar to the discussion made in the $^{77}$Se NMR study of FeSe \cite{baek15,baek16}, it is natural to consider that the local spin susceptibility, $A_\text{hf}\chi_\text{spin}$ in Eq. (\ref{knight}) is mainly responsible for the $C_4$ symmetry breaking at the As sites, proving that the nematic transition is electronically driven \footnote{Nonetheless,  in the nematic phase the asymmetry parameter 
	$\eta$ and thus the quadrupole term given by Eq. (\ref{quad}), \textit{by definition},  
	become finite regardless of the origin of nematicity. In particular, orbital order can strongly influence the  
	EFG and induce a large $\eta$ \cite{iye15,zhou16}. 
	Indeed, the finite quadrupole effect accounts for why the line splitting is not ascribed to the spin   
	contribution alone, i.e., it does not increase linearly with increasing $H$ in Fig. \ref{spec_comp}. 
	Besides, the quadrupole effect also accounts for the faster decrease of the average NMR shift of the split lines below \tnem\ [symbol $\times$ in Fig. \ref{fig_knight}(a)] than the unsplit line above \tnem, because it effects negative shifts of NMR spectra for any direction of the in-plane field \cite{bennet}.}.

\begin{figure}
	\centering
	\includegraphics[width=0.75\linewidth]{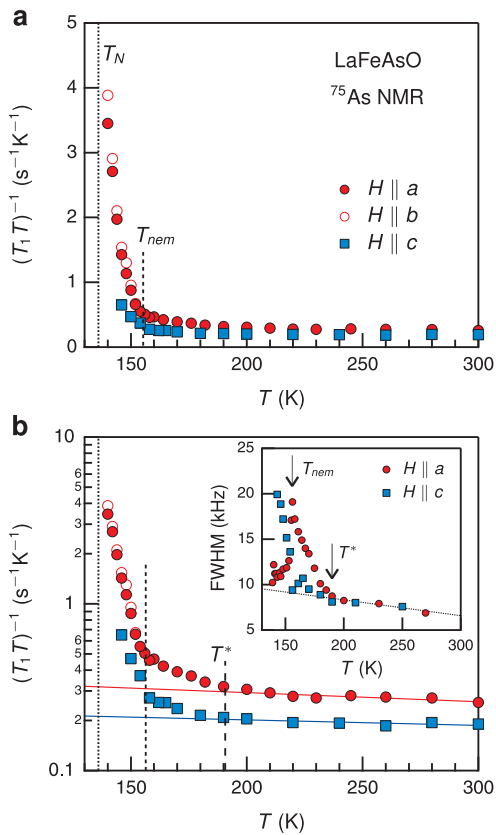}
	\caption{
		(a) Temperature dependence of 
		\as\ spin-lattice relaxation rate divided by $T$, \slrt.  
		(b) The same data as in (a) in a semi-log scale.  
		\slrt\ reveals a clear kink at \tnem\ before diverging at $T_N$, indicating  
		that SFs are abruptly enhanced by nematic order. The inset shows temperature dependence of FWHM. The FWHM for $H\parallel a$ starts to increase below a characteristic the temperature  
		$T^*>T_\text{nem}$ with respect to the background. In contrast, the data for $H\parallel c$ 
		does not change through $T^*$. } 
	\label{fig_t1t}
\end{figure}

Having established that the lattice 
distortion is not a primary order parameter for nematicity,
now we discuss the possible role of the spin degree of freedom for the nematic transition. For this purpose, we measured the spin-lattice relaxation rate \slr\ as a function of temperature, as the quantity 
\slrt\ 
probes SFs averaged over the Brillouin zone at very low energy.
The results are shown in Fig.  \ref{fig_t1t}(a). \slrt\ is nearly constant for both field orientations with a 
weak anisotropy at high temperatures.  At near $T_\text{nem}\sim 156$ K, however, it starts to upturn accompanying a strong anisotropy and diverges at $T_N$ for $H\parallel a$, being consistent with previous NMR studies \cite{fu12,hess17}. The strong development of SFs in the nematic phase raises the question whether SFs drives nematicity or it is a consequence of nematic ordering. 
In order to answer the question, we measured \slrt\ very carefully near \tnem\ for both $H\parallel a$ and  $H\parallel c$.  
Remarkably, we observed a sharp kink of \slrt\ exactly at \tnem\ for both field 
orientations, which is better shown in a semilog plot [Fig. \ref{fig_t1t}(b)]. This observation indicates that nematic ordering drastically enhances SFs in LaFeAsO, which is consistent with inelastic neutron scattering results \cite{zhang15a}. 
This can be qualitatively understood in terms of the dynamical spin susceptibility $\chi_\mathbf{q} \propto (r+ q^2)^{-1} $  where $r\sim \xi^{-2}$ with the magnetic correlation length $\xi$. As the nematic order parameter $\phi$ becomes nonzero below \tnem, $\xi$ is renormalized as $\xi^{-2} \rightarrow \xi^{-2} - \phi $ \cite{fernandes12} and thus $\chi_{\mathbf{q}}$  or  \slrt\ is strongly enhanced by the onset of nematic order. 

Interestingly, Fig. \ref{fig_t1t}(b) reveals that \slrt\ deviates from the 
background (solid lines) at a much higher temperature $T^*\sim 190$ K than \tnem. 
Moreover, we find that the FWHM of the \as\ line for 
$H\parallel a$ abruptly increases with respect to the background at $T^*$, as shown in the inset of Fig. 
\ref{fig_t1t}(b). 
These findings are similar to those  
observed in NaFe$_{1-x}$Co$_x$As \cite{zhou16}, where the \as\ line  
broadening with non-zero $\eta$ sets in at a temperature far above \tnem. Zhou 
\textit{et al} ascribed the 
development of the non-zero $\eta$ above \tnem\ to an 
incommensurate orbital order in the tetragonal phase. An 
alternative explanation may be the occurrence of the lattice 
softening due to nematic fluctuations \cite{fernandes10a}, which may lead to the 
local formation of nematic domains.
Regardless of its origin, our data indicate that the weak enhancement of SFs below $T^*$ is associated with nematic domains locally generated in some regions of the sample. 

In LaFeAsO, a magnetic transition takes place subsequent to a nematic 
transition, which differs from FeSe where superconducting order develops at a lower 
temperature without magnetic ordering. 
Recently, Chubukov et al.~\cite{chubukov16} argued that the hierarchy 
of orbital, magnetic, and superconducting instabilities is essentially determined by the 
largest Fermi energy $E_F$.  Namely, for a sufficiently small  
$E_F$, the leading instability is towards orbital order as in FeSe, while it is towards a 
spin density wave (SDW) or superconductivity as in other FeSCs for a large $E_F$. 
Despite the large difference of $E_F$ between LaFeAsO and FeSe, however, our NMR results in LaFeAsO and the comparison 
with those in FeSe \cite{baek15,bohmer15,baek16}  
strongly suggest that nematic order promotes SFs in the same way for both systems.  It should be noted that a sharp enhancement of SFs below \tnem\ is also observed in another magnetically ordered system, NaFeAs \cite{kitagawa11, ma11a}. That is, the proximity of a nematic state to long range magnetic order does not necessarily indicate that the spin degree of freedom is the driving force for the nematic transition in FeSCs \cite{lee12b,stanev13}. Rather, our NMR findings suggest that nematicity is generally driven by orbital order even in magnetically ordered FeSCs, although it can enhance a magnetic instability. 
However, this relationship between nematic order and magnetism cannot be established in the BaFe$_2$As$_2$-type system where magnetic and structural transitions occur simultaneously so that it is not possible to disentangle the orbital and spin degrees of freedom. 

In conclusion, by means of \as\ NMR, we have investigated the nematic and magnetic properties in high quality LaFeAsO single crystals. A sharp \as\ line splitting observed for $H\parallel a$ below \tnem\ has been proven to arise from the electronically driven twinned nematic domains. The \as\ spin-lattice relaxation data reveal that spin fluctuations are sharply enhanced by nematic order, similar to the behavior observed in non-magnetically ordered FeSe. We conclude that the leading instability for nematicity is identical for both FeSe and LaFeAsO, irrespective of whether long-range magnetic ordering occurs in the nematic state. 

\begin{acknowledgements}
This work has been supported by the Deutsche Forschungsgemeinschaft (Germany) 
via DFG Research Grant BA 4927/2-1. The work at POSTECH was supported by SRC Center for Topological Matter (No. 2011-0030785) and the Max Planck POSTECH/KOREA Research Initiative Program (No. 2016K1A4A4A01922028) through NRF in Korea.
\end{acknowledgements}

\bibliography{mybib.bib}





\end{document}